\documentclass{article}
\usepackage{ijcai19}
\usepackage[english]{babel}

\usepackage{hyperref}
\usepackage{times}
\usepackage{soul}
\usepackage{url}
\usepackage[utf8]{inputenc}
\usepackage{caption}
\usepackage{graphicx}
\usepackage{amsmath}
\usepackage{booktabs}
\usepackage{graphicx}
\usepackage{subcaption}
\title{Deep Learning Models for Calculation of Cardiothoracic Ratio from Chest Radiographs for Assisted Diagnosis of Cardiomegaly}

\author{
Tanveer Gupte\and
Mrunmai Niljikar\and
Manish Gawali\and
\\
Viraj Kulkarni\and
Amit Kharat\and
Aniruddha Pant
\affiliations
DeepTek Inc\\
\emails
}

\begin{document}

\maketitle

\begin{abstract}
We propose an automated method based on deep learning to compute the cardiothoracic ratio and detect the presence of cardiomegaly from chest radiographs. We develop two separate models to demarcate the heart and chest regions in an X-ray image using bounding boxes and use their outputs to calculate the cardiothoracic ratio. We obtain a sensitivity of 0.96 at a specificity of 0.81 with a mean absolute error of 0.0209 on a held-out test dataset and a sensitivity of 0.84 at a specificity of 0.97 with a mean absolute error of 0.018 on an independent dataset from a different hospital. We also compare three different segmentation model architectures for the proposed method and observe that Attention U-Net yields better results than SE-Resnext U-Net and EfficientNet U-Net. By providing a numeric measurement of the cardiothoracic ratio, we hope to mitigate human subjectivity arising out of visual assessment in the detection of cardiomegaly.

\end{abstract}

\section{Introduction}

Chest X-rays (CXR) are most commonly used for the diagnosis of heart and chest related pathologies. The research in computer-aided diagnosis of pathology from chest X-ray has progressed rapidly in the past few years, which has resulted in the generation of a massive corpus of open-source X-ray data along with ground truths, and the development of a large number of complex deep learning algorithms. However, the performance and outputs of complex deep learning algorithms on these large datasets are often subjective and lack an intuitive interpretable understanding of pathology.

Cardiomegaly manifests in the form of enlargement of the heart due to congenital causes, high blood pressure, or as a result of other pathologic conditions such as congenital heart diseases, valvular diseases, coronary artery disease, athletic heart, etc.  It presents itself with several forms of primary or acquired cardiomyopathies and may involve enlargement of the right, left, or both ventricles or the atria\cite{amin_siddiqui_2020}.
Diagnosis of cardiomegaly is primarily made using imaging techniques, which aid in measuring the heart’s size. Additionally, a quantitative measure called Cardiothoracic ratio (CTR) is also used to determine the presence of cardiomegaly. CTR is the ratio of maximal horizontal cardiac diameter (Wh) to maximal horizontal thoracic (Wt) diameter (inner edge of ribs/edge of pleura) as shown in figure 1.

\[ CTR = Wh / Wt \]

The range of CTR values between 0.42 and 0.50 indicates a normal condition. Values lesser than 0.42 or greater than 0.5 imply pathologic conditions. A higher CTR is suggestive of cardiomegaly.

\begin{figure}
\centering
\includegraphics[width=0.73\linewidth]{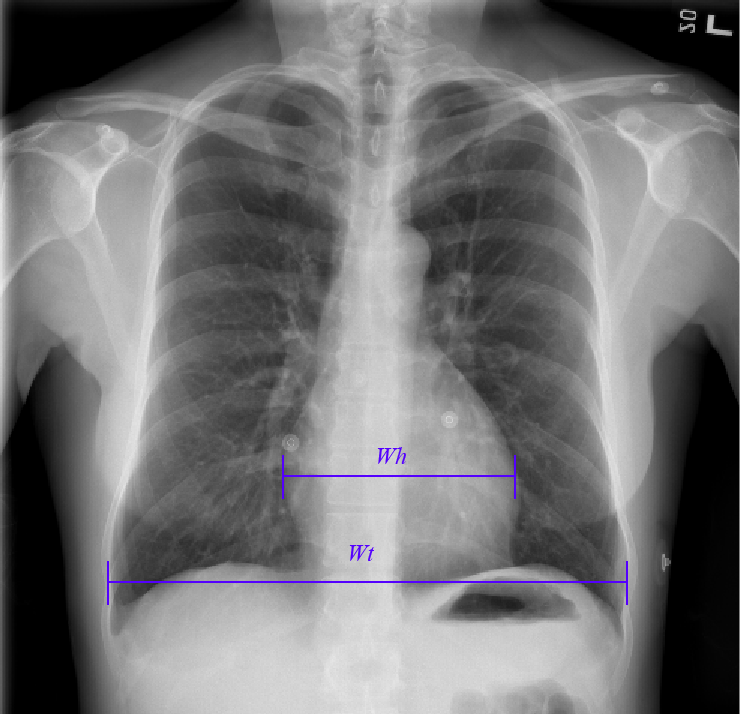}
\caption{Width of heart and thorax used for calculation of CTR}
\end{figure}

We propose an automated method that employs a U-Net-based architecture to segment the lungs and the heart from a CXR followed by calculation of the cardiothoracic ratio to determine the presence or absence of cardiomegaly. The segmentation outputs for the lung and heart region and the automated CTR calculation makes the overall decision given by the deep learning model interpretable. Moreover, we conduct three experiments with three different segmentation-based architectures to decide the best segmentation-based architecture compatible with the proposed method.  

\section{Related Work}

Computer aided-diagnosis has been used to detect cardiomegaly in a chest X-ray with or without using deep learning techniques. Initially, cardiomegaly was identified by calculating CTR using the segmentation method\cite{2017SPIE10134E..0KD}\cite{hasan_lee_kim_lim_2012}\cite{ginneken_stegmann_loog_2006}. In one such study, Candemir et al.\cite{candemir2016automatic} used pre-segmented images to localize the heart and lungs on the CXRs to extract the detailed radiographic index from the heart and lung boundaries. These radiographic indices assisted in the diagnosis of cardiomegaly. The study used multiple radiographic indexes to build a classifier that classified patients with cardiomegaly with a sensitivity of 0.77 at a specificity of 0.76.

Islam et al.\cite{islam2017abnormality} and Candemir et al.\cite{candemir2018deep} proposed the use of simple classification deep learning architectures to detect cardiomegaly. To predict Cardiomegaly’s presence with more objectivity, CTR had to be calculated using deep learning techniques. A fully convolutional neural network was employed to segment CXR images and calculate CTR. Li et al.\cite{8675927} adopted a UNET architecture\cite{ronneberger2015u} that identified and localized the lungs and heart to get the heart and chest width, which were then used to calculate CTR. They used a private dataset of 5000 posteroanterior (PA) view CXRs, in which the lungs and heart boundaries were annotated.

A similar technique was used by Chamveha et al.\cite{chamveha2020automated} where they used UNET architecture with VGG-16\cite{simonyan2014very} as the backbone of the encoder. Montgomery\cite{jaeger2014two}, JSRT\cite{shiraishi2000development}, and a subset of the NIH dataset\cite{wang2017chestx} were used for training the model and the validation set contained images from CheXpert dataset. Human radiologists evaluated the obtained CTR measurements, and 76.5\%  of the AI results were accepted and included in medical reports without any need for adjustment. Sogancioglu et al.\cite{sogancioglu2020cardiomegaly}  compared segmentation and classification approaches and demonstrated that segmentation models perform better for detecting cardiomegaly on chest radiographs. Moreover, they indicated that segmentation models needed fewer images to learn and the output is interpretable as compared to classification models.

\section{Data and Methods}
\subsection{Data}
We aggregated the chest X-ray images from an open-source research dataset NIH\cite{wang2017chestx},  two private hospitals (which we name as D1, D2), and a private dataset which was collected from population screening (which we name as D3). A team of expert radiologists annotated the CXRs by drawing bounding boxes around the heart and chest region using the VIA Annotation Software\cite{dutta2019via}. After drawing the bounding boxes, the CTR is automatically calculated by the software by considering the relative cardiac width and thoracic width from the annotated bounding boxes. A total of 2623 CXRs (table 1) were used for the study. For the NIH dataset, out of 30,805 patients with 112,120 radiographs, we randomly sampled 1440 X-rays of 1440 patients. The sampling was done so that cardiomegaly was present in half of the X-ray images (720) and not present in the other half (720). Both sets had 50\% CXRs with an anteroposterior (AP) view (360) and the other half with a PA view(360). The reason for this sampling was to create enough data variation to increase the robustness of the model. 1000 X-ray images from D1 and D3 were used. We used 183 images from D2 to evaluate the model’s performance on an out-of-source dataset. For training, 1952 CXRs were used. 244 CXRs were used for validation, and 244 CXRs were used for testing the result.

\begin{table}[hbt!]
\begin{center}
\begin{tabular}{|l|l|l|l|l|}
\hline
Dataset & Train & Validation & Test      & Total \\ \hline
NIH     & 1152  & 144        & 144       & 1440  \\ \hline
D1 + D3 & 800   & 100        & 100       & 1000  \\ \hline
D2      & -     & -          & 183       & 183   \\ \hline
Total   & 1952  & 244        & 244 + 183 & 2623  \\ \hline
\end{tabular}
\caption{Distribution of Cardiomegaly CXR's}
\end{center}
\end{table}

\subsection{Pre-processing and Augmentation}

\begin{figure*}
\includegraphics[width=\linewidth]{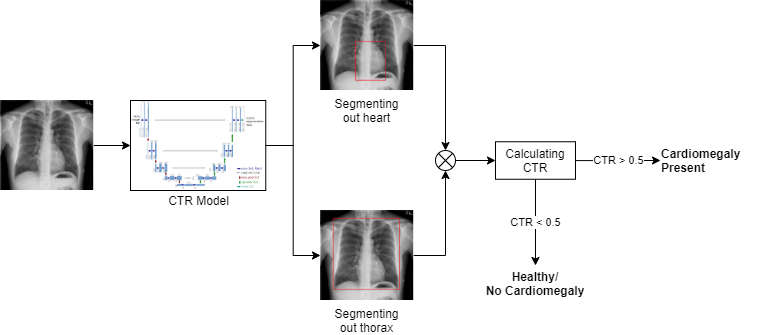}
\caption{Illustration of the architecture of the segmentation pipeline for calculation of CTR to determine the presence of Cardiomegaly. The model gives two segmentation maps for heart and thorax, which are then processed to calculate CTR.}
\end{figure*}

The CXRs of various dimensions were reduced to 512x512x3 pixels. All the CXRs were normalized before training such that values of pixels ranging from 0 to 255 were bounded in the range 0 to 1. The model was trained to get a dual-channel output. The first channel of the output was a heart’s mask and the second channel was the thorax’s mask. 

We experimented with different types of augmentations to upsample the training dataset\cite{hussain2017differential}. We resorted to geometric augmentations (except for Gaussian blur) of the CXRs as we wanted the model to detect the width of the heart and thorax efficiently. We found that combining shearing (along with the x and y-axis), scaling, a little bit of grid distortion, and gaussian blurring improved the model’s results. The training and validation dataset was randomly upsampled by 75\%. The augmentation was done so that only one or two of the augmentations, as mentioned earlier, were performed at a time. To maintain the output accuracy while augmentation, the output masks for the heart and thorax were also changed to match the correct outlines of the heart and thorax on augmented CXR. To avoid randomness while training every different model, the augmentations of the CXRs were done only once, and the subsequent data, along with output masks, were stored. The final training dataset had 1952 original CXRs and 1464 (75\% upsampled) augmented CXRs, i.e., a total of 3416 CXRs.

\subsection{Procedure and Post-processing}

We used the same training and validation datasets for training on all three models. Adam optimizer\cite{kingma2014adam} and binary cross-entropy loss were used with a learning rate that was reduced on the plateau i.e. the learning rate was reduced when the metric (validation loss) stopped improving. The model which had the lowest validation loss was saved and used for analysis. While inferring the model, for each CXR, the network predicts a region of interest with bounding coordinates for both heart and chest. These coordinates were used to determine the width of the chest and heart, which was then used to calculate the CTR. The ground truth was selected based on annotations, i.e., if the CTR for annotated CXRs was more than 0.5, it was labeled as Cardiomegaly. 

The output of the model contained a pixel-wise probability for the input image. The pixel probability was thresholded to get a clear-cut mask. The pixel probability of less than the threshold value was converted to 0, and anything above it was converted to 1. Following the thresholding, morphological transformations\cite{sreedhar2012enhancement}\cite{ravi2013morphological} were performed on the output masks to reduce the noise and error. The morphological transformation did not improve the specificity and the sensitivity of the model; however, it reduced the MAE and RMSE (Root Mean Squared Error). The transformations that yielded better results were erosion for two iterations, followed by dilation for one iteration.  The change in the output masks doesn’t seem apparent to human eyes, but we observed MAE reduction by 12.5\% and RMSE by 9\%.

The performance of all three models’ was calculated on two independent datasets. One was a held-out test dataset with the same source as the training and validation dataset, while the other was an out-of-source dataset D2. We compared our results for three different UNET \cite{ronneberger2015u} based architectures. The first was enhanced UNET with Spatial-attention gate\cite{oktay2018attention}\cite{khanh2020enhancing} (Attention UNET) with Xception encoder\cite{chollet2017xception}. The second one was UNET with Squeeze-and-Excitation\cite{hu2018squeeze} network blocks incorporated with the ResNext50 \cite{xie2017aggregated} (SE-ResNext 50) backbone. The third one was simple UNET with EfficientNet-b4\cite{tan2019efficientnet}\cite{9150621}  as its encoder. All of these networks were pre-trained with ImageNet\cite{5206848} weights.

\section{Results}

\begin{figure}[!ht]
\centering
\includegraphics[width=\linewidth]{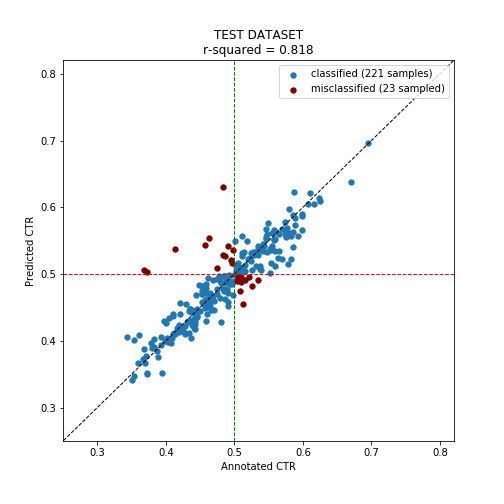}
\caption{Scatter plot of annotated CTR and predicted CTR on hold out test dataset}
\end{figure}

\begin{figure}[!ht]
\centering
\includegraphics[width=\linewidth]{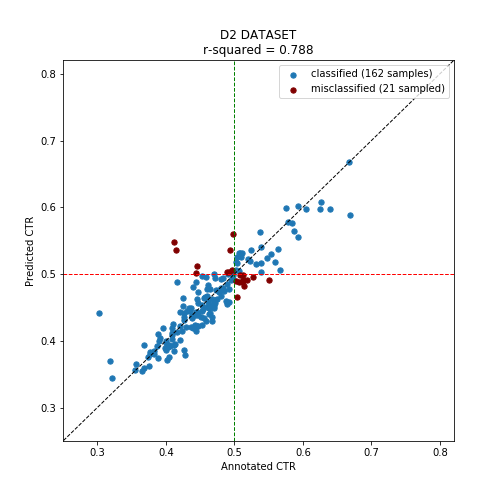}
\caption{Scatter plot of annotated CTR and predicted CTR on D2 test dataset}
\end{figure}

\begin{figure}
     \centering
     \begin{subfigure}{\linewidth}
     \centering
         \includegraphics[width=\linewidth]{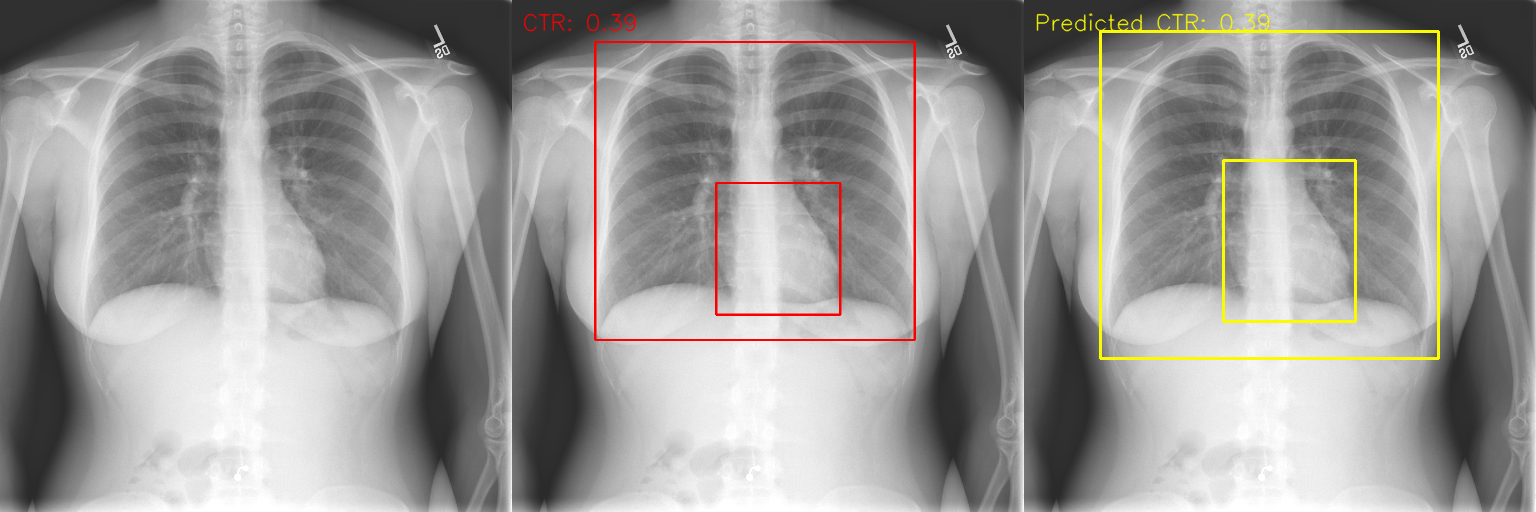}
         \caption{Radiograph with CTR 0.39; Cardiomegaly absent}
     \end{subfigure}
     \hfill
     \centering
     \begin{subfigure}{\linewidth}
     \centering
         \includegraphics[width=\linewidth]{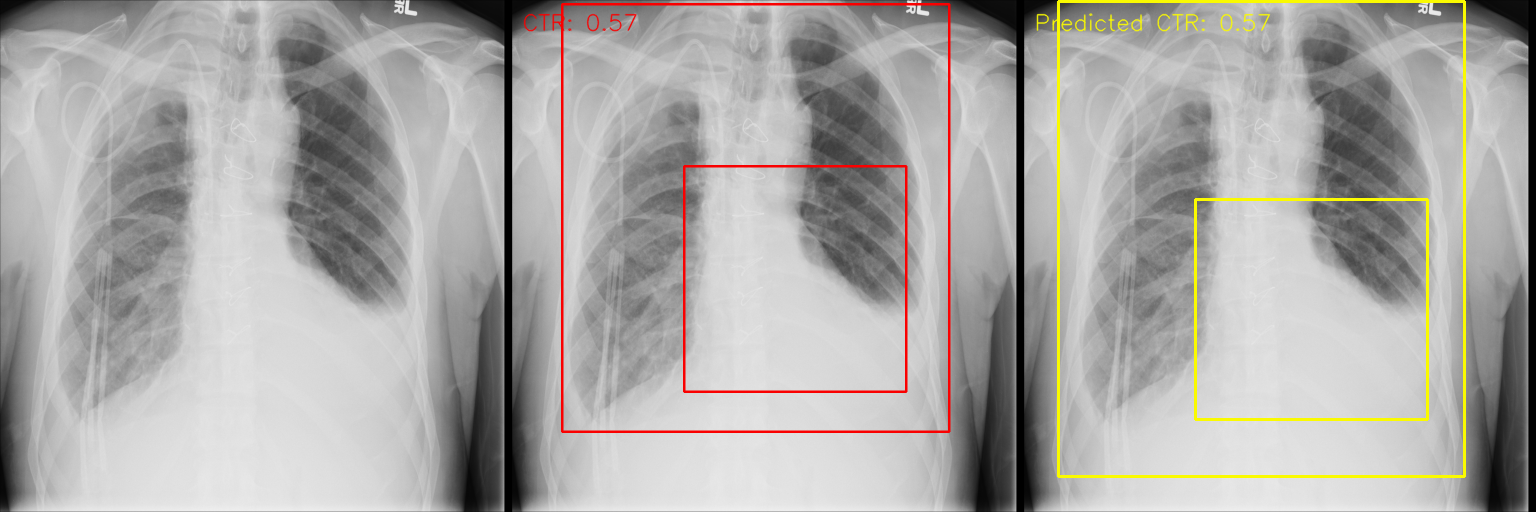}
         \caption{Radiograph with CTR 0.57; Cardiomegaly present}
     \end{subfigure}
     \caption{Visual presentations of the original, radiologist-annotated(red boxes) and AI model predicted (yellow boxes) radiographs.
}
\end{figure}   

Three different multi-class segmentation architectures were used to train three different models on the same datasets to determine the best performing robust model. Although inherently, all were UNET architectures, the encoders and the intermediate layers were different for each model. Hyperparameter optimization, image augmentation, and image processing were done to ensure the best possible results. The comparison of these models on the test dataset and D2 dataset can be seen in Tables 2 and 3, respectively. Since the results were close, it was hard to assess which model performs better for cardiomegaly classification based on CTR. It can be observed that the best performance classification metrics, i.e., sensitivity, specificity, and F-1 score, vary for all the models. Also, a trade-off between sensitivity and specificity cannot be altered as the cut-off for positive prediction of cardiomegaly is based on the theoretical cut-off value of CTR at 0.5. For classification performance, UNET SE-Resnext has the lowest performance for the D2 dataset, lowest F-1 score on the test dataset, making it the weakest contender. On the test dataset, the F-1 score of the UNET EfficientNet model is only marginally higher than Attention UNET, whereas the latter model has a significantly higher f-1 score on the D2 dataset. For regression metrics, however, the results were explicit. The Mean Absolute Error (MAE) and Root Mean Squared Error (RMSE) for Attention UNET are the lowest indicating far better masks resulting in much accurate CTR calculation. Attention UNET was used for further analysis due to its better performance. The scatter plot of predicted CTR and annotated CTR can be seen in figure 3 and 4. The misclassification of cardiomegaly, where CTR is close to 0.5, is inevitable due to inter-reader variability. The misclassification cases, where the difference between annotated and predicted CTR is significantly high, are 6 for test datasets and only 5-6 for D2 datasets.

\begin{table}[hbt!]
\begin{center}
\begin{tabular}{|l|l|l|l|}
\hline
Metrics     & \begin{tabular}[c]{@{}l@{}}Attention \\ UNET\end{tabular} & \begin{tabular}[c]{@{}l@{}}SE-Resnext\\ UNET\end{tabular} & \begin{tabular}[c]{@{}l@{}}Efficient Net\\ UNET\end{tabular} \\ \hline
Sensitivity & 0.96                                                      & 0.87                                                      & 0.94                                                         \\ \hline
Specificity & 0.81                                                      & 0.86                                                      & 0.83                                                         \\ \hline
F-1 Score   & 0.87                                                      & 0.86                                                      & 0.88                                                         \\ \hline
MAE         & 0.0209                                                    & 0.0206                                                    & 0.0328                                                       \\ \hline
RMSE        & 0.0312                                                    & 0.0317                                                    & 0.0798                                                       \\ \hline
\end{tabular}
\caption{Results on the held-out test dataset}
\end{center}
\end{table}

\begin{table}[hbt!]
\begin{center}
\begin{tabular}{|l|l|l|l|}
\hline
Metrics     & \begin{tabular}[c]{@{}l@{}}Attention \\ UNET\end{tabular} & \begin{tabular}[c]{@{}l@{}}SE-Resnext\\ UNET\end{tabular} & \begin{tabular}[c]{@{}l@{}}Efficient Net\\ UNET\end{tabular} \\ \hline
Sensitivity & 0.87                                                      & 0.8                                                       & 0.93                                                         \\ \hline
Specificity & 0.97                                                      & 0.91                                                      & 0.88                                                         \\ \hline
F-1 Score   & 0.88                                                      & 0.78                                                      & 0.81                                                         \\ \hline
MAE         & 0.0181                                                    & 0.0206                                                    & 0.0248                                                       \\ \hline
RMSE        & 0.0282                                                    & 0.0317                                                    & 0.058                                                        \\ \hline
\end{tabular}
\caption{ Results on D2 (out-of-source) dataset}
\end{center}
\end{table}

\section{Conclusion}

The diagnosis of cardiomegaly is subjective and varies from radiologist-to-radiologist. A limitation of the classification approach to detect cardiomegaly is that the results are not interpretable and lack objectivity.  Furthermore, many radiologists diagnose cardiomegaly only if it’s severe, and many borderline cases go unnoticed\cite{olatunji2019caveats}. Using the classification approach will not improve the correctness of the diagnosis as it does not CTR into account. In this study, we built a model to detect changes in the width of the cardiac silhouette. Unlike other pathologies, cardiomegaly is a quantifiable pathology, and calculating the CTR will help establish the necessary output. Our adapted Attention UNET architecture exhibited excellent chest X-ray image segmentation.

To determine cardiomegaly, CTR is usually calculated on the posteroanterior (PA) view of the CXR rather than anteroposterior due to better visualization of the cardiac silhouette in PA view\cite{chon2011calculation}. We deliberately chose to use images without taking into account whether they are AP or PA views. This step was done to correctly identify the cardiac silhouette and thoracic outline to get the width of both, respectively. The ratio calculated will guide the radiologists in clinical settings without subjecting them into considering the established ideology of a better CTR assessment in PA view.

Another challenge in our study was to make the model robust as to make it work in practical clinical settings. Since different hospitals use different X-ray machines with various manual and automatic settings before exposing the radiographic films, the seemingly similar CXRs may show irregularities. Thus a model trained on one-source of data usually performs poorly on another source of data. We tackled this problem by incorporating data from multiple sources while training and using image augmentations. A dataset from an entirely different hospital setup was kept aside to evaluate out-of-source performance. The model's ability to perform well on out-of-source hospital dataset was demonstrated.

\bibliographystyle{ieeetr}
\nocite{*}
\bibliography{bibliography}

\begin{thebibliography}{10}

\bibitem{amin_siddiqui_2020}
H.~Amin and W.~J. Siddiqui, ``Cardiomegaly,'' Nov 2020.

\bibitem{2017SPIE10134E..0KD}
A.~H. {Dallal}, C.~{Agarwal}, M.~R. {Arbabshirani}, A.~{Patel}, and G.~{Moore},
  ``{Automatic estimation of heart boundaries and cardiothoracic ratio from
  chest x-ray images},'' in {\em Society of Photo-Optical Instrumentation
  Engineers (SPIE) Conference Series}, vol.~10134 of {\em Society of
  Photo-Optical Instrumentation Engineers (SPIE) Conference Series},
  p.~101340K, Mar. 2017.

\bibitem{hasan_lee_kim_lim_2012}
M.~A. Hasan, S.-L. Lee, D.-H. Kim, and M.-K. Lim, ``Automatic evaluation of
  cardiac hypertrophy using cardiothoracic area ratio in chest radiograph
  images,'' {\em Computer Methods and Programs in Biomedicine}, vol.~105,
  no.~2, p.~95–108, 2012.

\bibitem{ginneken_stegmann_loog_2006}
B.~V. Ginneken, M.~B. Stegmann, and M.~Loog, ``Segmentation of anatomical
  structures in chest radiographs using supervised methods: a comparative study
  on a public database,'' {\em Medical Image Analysis}, vol.~10, no.~1,
  p.~19–40, 2006.

\bibitem{candemir2016automatic}
S.~Candemir, S.~Jaeger, W.~Lin, Z.~Xue, S.~Antani, and G.~Thoma, ``Automatic
  heart localization and radiographic index computation in chest x-rays,'' in
  {\em Medical Imaging 2016: Computer-Aided Diagnosis}, vol.~9785, p.~978517,
  International Society for Optics and Photonics, 2016.

\bibitem{islam2017abnormality}
M.~T. Islam, M.~A. Aowal, A.~T. Minhaz, and K.~Ashraf, ``Abnormality detection
  and localization in chest x-rays using deep convolutional neural networks,''
  {\em arXiv preprint arXiv:1705.09850}, 2017.

\bibitem{candemir2018deep}
S.~Candemir, S.~Rajaraman, G.~Thoma, and S.~Antani, ``Deep learning for grading
  cardiomegaly severity in chest x-rays: an investigation,'' in {\em 2018 IEEE
  Life Sciences Conference (LSC)}, pp.~109--113, IEEE, 2018.

\bibitem{8675927}
Z.~{Li}, Z.~{Hou}, C.~{Chen}, Z.~{Hao}, Y.~{An}, S.~{Liang}, and B.~{Lu},
  ``Automatic cardiothoracic ratio calculation with deep learning,'' {\em IEEE
  Access}, vol.~7, pp.~37749--37756, 2019.

\bibitem{ronneberger2015u}
O.~Ronneberger, P.~Fischer, and T.~Brox, ``U-net: Convolutional networks for
  biomedical image segmentation,'' in {\em International Conference on Medical
  image computing and computer-assisted intervention}, pp.~234--241, Springer,
  2015.

\bibitem{chamveha2020automated}
I.~Chamveha, T.~Promwiset, T.~Tongdee, P.~Saiviroonporn, and
  W.~Chaisangmongkon, ``Automated cardiothoracic ratio calculation and
  cardiomegaly detection using deep learning approach,'' {\em arXiv preprint
  arXiv:2002.07468}, 2020.

\bibitem{simonyan2014very}
K.~Simonyan and A.~Zisserman, ``Very deep convolutional networks for
  large-scale image recognition,'' {\em arXiv preprint arXiv:1409.1556}, 2014.

\bibitem{jaeger2014two}
S.~Jaeger, S.~Candemir, S.~Antani, Y.-X.~J. W{\'a}ng, P.-X. Lu, and G.~Thoma,
  ``Two public chest x-ray datasets for computer-aided screening of pulmonary
  diseases,'' {\em Quantitative imaging in medicine and surgery}, vol.~4,
  no.~6, p.~475, 2014.

\bibitem{shiraishi2000development}
J.~Shiraishi, S.~Katsuragawa, J.~Ikezoe, T.~Matsumoto, T.~Kobayashi, K.-i.
  Komatsu, M.~Matsui, H.~Fujita, Y.~Kodera, and K.~Doi, ``Development of a
  digital image database for chest radiographs with and without a lung nodule:
  receiver operating characteristic analysis of radiologists' detection of
  pulmonary nodules,'' {\em American Journal of Roentgenology}, vol.~174,
  no.~1, pp.~71--74, 2000.

\bibitem{wang2017chestx}
X.~Wang, Y.~Peng, L.~Lu, Z.~Lu, M.~Bagheri, and R.~M. Summers, ``Chestx-ray8:
  Hospital-scale chest x-ray database and benchmarks on weakly-supervised
  classification and localization of common thorax diseases,'' in {\em
  Proceedings of the IEEE conference on computer vision and pattern
  recognition}, pp.~2097--2106, 2017.

\bibitem{sogancioglu2020cardiomegaly}
E.~Sogancioglu, K.~Murphy, E.~Calli, E.~T. Scholten, S.~Schalekamp, and
  B.~Van~Ginneken, ``Cardiomegaly detection on chest radiographs: Segmentation
  versus classification,'' {\em IEEE Access}, vol.~8, pp.~94631--94642, 2020.

\bibitem{dutta2019via}
A.~Dutta and A.~Zisserman, ``The via annotation software for images, audio and
  video,'' in {\em Proceedings of the 27th ACM International Conference on
  Multimedia}, pp.~2276--2279, 2019.

\bibitem{hussain2017differential}
Z.~Hussain, F.~Gimenez, D.~Yi, and D.~Rubin, ``Differential data augmentation
  techniques for medical imaging classification tasks,'' in {\em AMIA Annual
  Symposium Proceedings}, vol.~2017, p.~979, American Medical Informatics
  Association, 2017.

\bibitem{kingma2014adam}
D.~P. Kingma and J.~Ba, ``Adam: A method for stochastic optimization,'' {\em
  arXiv preprint arXiv:1412.6980}, 2014.

\bibitem{sreedhar2012enhancement}
K.~Sreedhar and B.~Panlal, ``Enhancement of images using morphological
  transformation,'' {\em arXiv preprint arXiv:1203.2514}, 2012.

\bibitem{ravi2013morphological}
S.~Ravi and A.~Khan, ``Morphological operations for image processing:
  understanding and its applications,'' in {\em Proc. 2nd National Conference
  on VLSI, Signal processing \& Communications NCVSComs-2013}, 2013.

\bibitem{oktay2018attention}
O.~Oktay, J.~Schlemper, L.~L. Folgoc, M.~Lee, M.~Heinrich, K.~Misawa, K.~Mori,
  S.~McDonagh, N.~Y. Hammerla, B.~Kainz, {\em et~al.}, ``Attention u-net:
  Learning where to look for the pancreas,'' {\em arXiv preprint
  arXiv:1804.03999}, 2018.

\bibitem{khanh2020enhancing}
T.~L.~B. Khanh, D.-P. Dao, N.-H. Ho, H.-J. Yang, E.-T. Baek, G.~Lee, S.-H. Kim,
  and S.~B. Yoo, ``Enhancing u-net with spatial-channel attention gate for
  abnormal tissue segmentation in medical imaging,'' {\em Applied Sciences},
  vol.~10, no.~17, p.~5729, 2020.

\bibitem{chollet2017xception}
F.~Chollet, ``Xception: Deep learning with depthwise separable convolutions,''
  in {\em Proceedings of the IEEE conference on computer vision and pattern
  recognition}, pp.~1251--1258, 2017.

\bibitem{hu2018squeeze}
J.~Hu, L.~Shen, and G.~Sun, ``Squeeze-and-excitation networks,'' in {\em
  Proceedings of the IEEE conference on computer vision and pattern
  recognition}, pp.~7132--7141, 2018.

\bibitem{xie2017aggregated}
S.~Xie, R.~Girshick, P.~Doll{\'a}r, Z.~Tu, and K.~He, ``Aggregated residual
  transformations for deep neural networks,'' in {\em Proceedings of the IEEE
  conference on computer vision and pattern recognition}, pp.~1492--1500, 2017.

\bibitem{tan2019efficientnet}
M.~Tan and Q.~V. Le, ``Efficientnet: Rethinking model scaling for convolutional
  neural networks,'' {\em arXiv preprint arXiv:1905.11946}, 2019.

\bibitem{9150621}
B.~{Baheti}, S.~{Innani}, S.~{Gajre}, and S.~{Talbar}, ``Eff-unet: A novel
  architecture for semantic segmentation in unstructured environment,'' in {\em
  2020 IEEE/CVF Conference on Computer Vision and Pattern Recognition Workshops
  (CVPRW)}, pp.~1473--1481, 2020.

\bibitem{5206848}
J.~{Deng}, W.~{Dong}, R.~{Socher}, L.~{Li}, {Kai Li}, and {Li Fei-Fei},
  ``Imagenet: A large-scale hierarchical image database,'' in {\em 2009 IEEE
  Conference on Computer Vision and Pattern Recognition}, pp.~248--255, 2009.

\bibitem{olatunji2019caveats}
T.~Olatunji, L.~Yao, B.~Covington, A.~Rhodes, and A.~Upton, ``Caveats in
  generating medical imaging labels from radiology reports,'' {\em arXiv
  preprint arXiv:1905.02283}, 2019.

\bibitem{chon2011calculation}
S.~B. Chon, W.~S. Oh, J.~H. Cho, S.~S. Kim, and S.-J. Lee, ``Calculation of the
  cardiothoracic ratio from portable anteroposterior chest radiography,'' {\em
  Journal of Korean Medical Science}, vol.~26, no.~11, pp.~1446--1453, 2011.

\end{thebibliography}
\end{document}